# One-dimensional van der Waals heterojunction diode

Ya Feng,[1*] Henan Li,[1] Taiki Inoue,[1,2] Shohei Chiashi,[1] Slava V. Rotkin,[3] Rong Xiang,[1] Shigeo Maruyama[1*]

[1]Department of Mechanical Engineering, School of Engineering, The University of Tokyo, Tokyo 113-8656, Japan

[2]Department of Applied Physics, Graduate School of Engineering, Osaka University, Osaka 565-0871, Japan

[3]Department of Engineering Science and Mechanics, Materials Research Institute, The Pennsylvania State University, Millennium Science Complex, University Park, PA 16802, USA

TOC

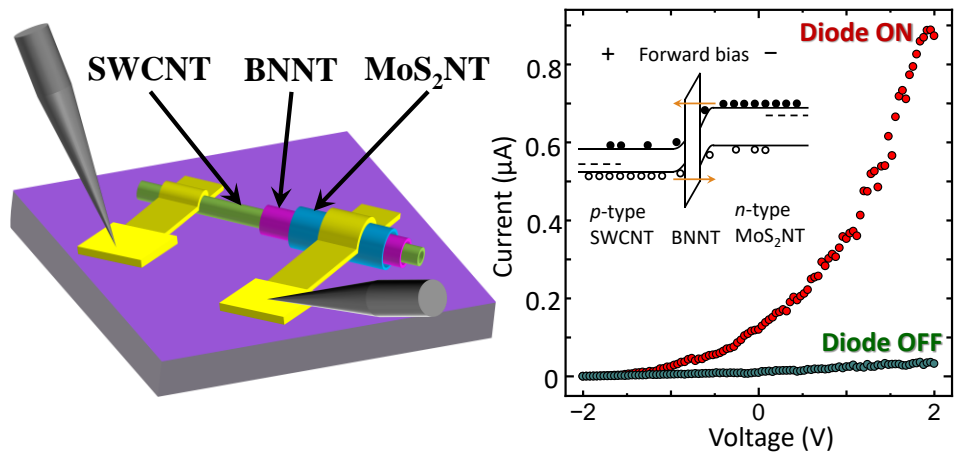

ABSTRACT

The synthesis of one-dimensional van der Waals heterostructures was realized recently, which opens up new possibilities for prospective applications in electronics and optoelectronics. The even reduced dimension will enable novel properties and further miniaturization beyond the capabilities of its two-dimensional counterparts have revealed. The natural doping results in *p*-type electrical characteristics for semiconducting single-walled carbon nanotubes, while *n*-type for molybdenum disulfide with conventional noble metal contacts. Therefore, we demonstrate here a one-dimensional heterostructure nanotube of 11-nm-wide, with the coaxial assembly of semiconducting single-walled carbon nanotube, insulating boron nitride nanotube, and semiconducting molybdenum disulfide nanotube which induces a radial semiconductor-insulator-semiconductor heterojunction. When opposite potential polarity was applied on semiconducting single-walled carbon nanotube and molybdenum disulfide nanotube, respectively, the rectifying effect was materialized.

KEYWORDS single-walled carbon nanotubes, one-dimensional heterostructure, heterojunction diode, boron nitride nanotubes, molybdenum disulfide, van der Waals

INTRODUCTION

Semiconductor *p-n* junctions are fundamental to build up state of art optoelectronic architectures[1]. The emerging two-dimensional (2D) van der Waals assemblies[2, 3], including atomically thin semiconducting transition metal dichalcogenides (TMD) and engineered graphene have pushed *p-n* junctions to an ultimate thickness limit, which enable tunneling diodes with negative differential resistance (NDR)[4], tunneling transistors[5, 6], novel photovoltaic devices[7, 8], and quantum wells light-emitting diodes (LED)[9]. With further dimension confining, the rolled-up graphene, one-dimensional (1D) single-walled carbon nanotube (SWCNT) can induce a direct bandgap, therefore, allow various applications in optoelectronics[10-13]. Efforts have been made to realize a single SWCNT diode, through chemical doping[14, 15] which suffers from short durability, introducing asymmetric metal contacts[16] which involves multiple intricate metal deposition processes, or electrostatic doping[17, 18] which, however, is limited by the gate leakage and also requires processing with multiple metals[19], in order to generate a built-in potential to drive flow of carriers unidirectionally.

The remarkable performance improvement of 2D electronic devices by van der Waals layers stacking[20, 21], as well as the versatilities it presents[4, 9] has intrigued the exploration of van der Waals heterostructures in 1D field, and such an 1D counterpart templating with SWCNT has been synthesized by chemical vapor deposition (CVD) recently[22]. Therefore, an 1D ultra-thin heterojunction can be expected from naturally *p*-doped semiconducting SWCNT[23] and *n*-doped molybdenum disulfide nanotube[24] ($MoS_2NT$) heterostructure. In the present work, we propose a radial semiconductor-insulator-semiconductor (S-I-S) heterojunction with 1D heterostructure composed of coaxial SWCNT, boron nitride nanotube (BNNT), and $MoS_2NT$. In contrast to lateral 1D devices that suffer from fringe fields and incomplete electrostatic gating, radial (wrap-around) geometry gives the ultimate control of 1D charge density[25, 26]. We synthesized micrometer-long SWCNT bridging over silicon poles, then coated it with BNNT to increase the diameter in order to complete the ultimate coating of $MoS_2NT$. This novel 1D S-I-S heterojunction results in significant rectifying effect with one electrode touching inner semiconducting SWCNT while the other in contact with outside $MoS_2NT$. One-dimensional S-I-S heterojunction diode presented in this work as scalable as single molecule diodes[27, 28], can readily fit into current semiconducting industry, providing an alternative method to further miniaturize optoelectronic building blocks. The feasibility shown by the current 1D heterojunction diode and the multifunctionalities inherent in heterostructure imply the tremendous potential in near future electronic and optoelectronic applications.

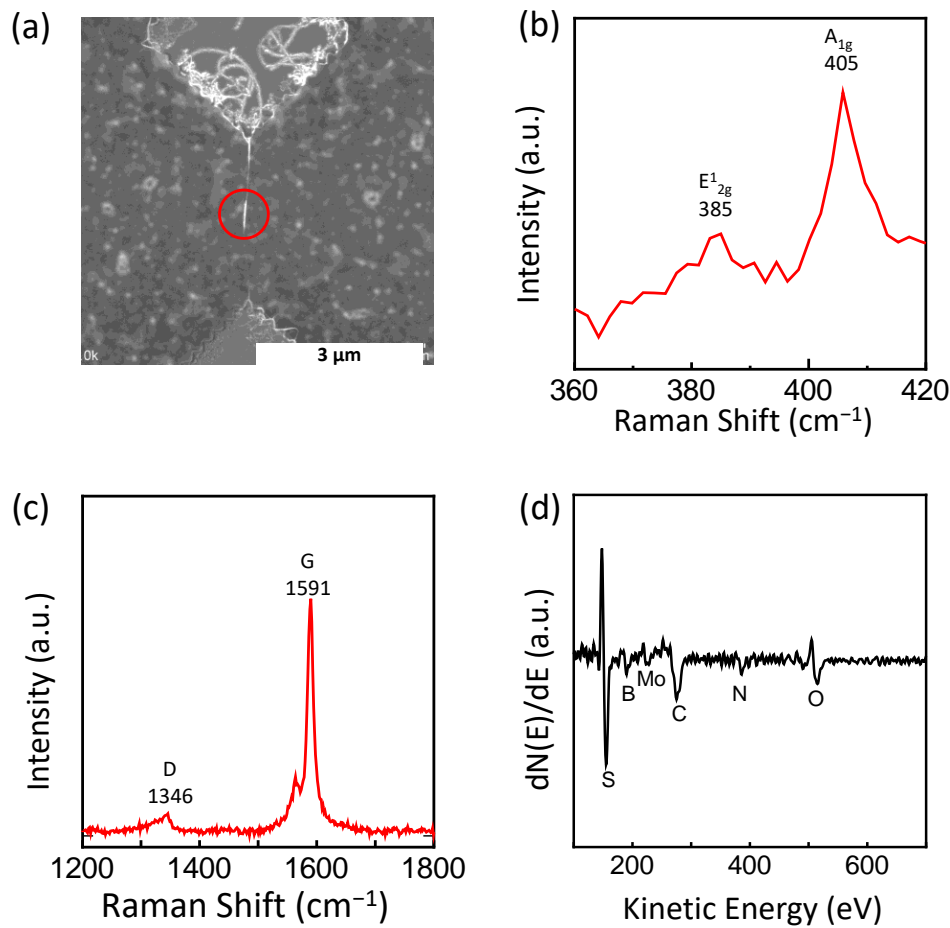

**Figure 1**. Characterizations of the suspended SWCNT/BNNT/MoS$_2$NT heterostructure nanotube. (a) SEM image of the suspended nanotube. Raman spectrum (the laser spot was focused on the red-circle area in (a)) shows the fingerprint peaks from MoS$_2$ in (b), as well as D and G peaks from SWCNT in (c). (d) AES spectrum from the suspended nanotube area.

RESULTS AND DISCUSSIONS

The micrometer-long suspended SWCNT was prepared over Si poles with confined Co catalysts on the top as illustrated in Fig. S1. The small patterned catalyst areas effectively avoided SWCNT bundles during growth. Thereafter, BNNT and MoS$_2$NT coating were sequentially conducted. Non-uniformities accumulated along suspended nanotubes after BNNT coating because of the rather randomly distributed nucleation sites. Further MoS$_2$NT coating along the suspended nanotube presents sharp contrasts as shown in the scanning electron microscope (SEM) images of Fig. S2. Confocal Raman spectrum with laser wavelength of 532 nm focusing on the red-circle site in Fig. 1(a), which is seemingly thicker and brighter under SEM, exhibits the fingerprint peaks of molybdenum disulfide (MoS$_2$) locate at 385 cm$^{-1}$ and 405 cm$^{-1}$ Raman shifts in Fig. 1(b), respectively, as well as D and G peaks from SWCNT shown in Fig. 1(c). The weak nonresonant boron nitride (BN) Raman scattering that overlaps with D peak of CNT[29] is unable to manifest here. The full spectrum including radial breathing mode (RBM) from SWCNT as well as arrow-marked peaks from the substrate is presented in Fig. S3. Auger electron spectroscopy (AES) was employed to verify the composition of the suspended heterostructure nanotube. Figure 1(d) demonstrates AES spectrum from as-grown heterostructure nanotube with excitation electron beam of 10 kV and 10 nA, which detected carbon, boron, nitrogen, molybdenum, and sulfur atoms from the suspended nanotube as shown in Fig. 1(a), where oxygen is originated from environmental adsorption or substrate. From the above analysis, we can draw a conclusion that the relatively brighter parts in SEM images surfaced after the second and third CVD processes on the suspended heterostructure nanotube can be attributed to the successful coating of BNNT and/or MoS$_2$NT van der Waals layers.

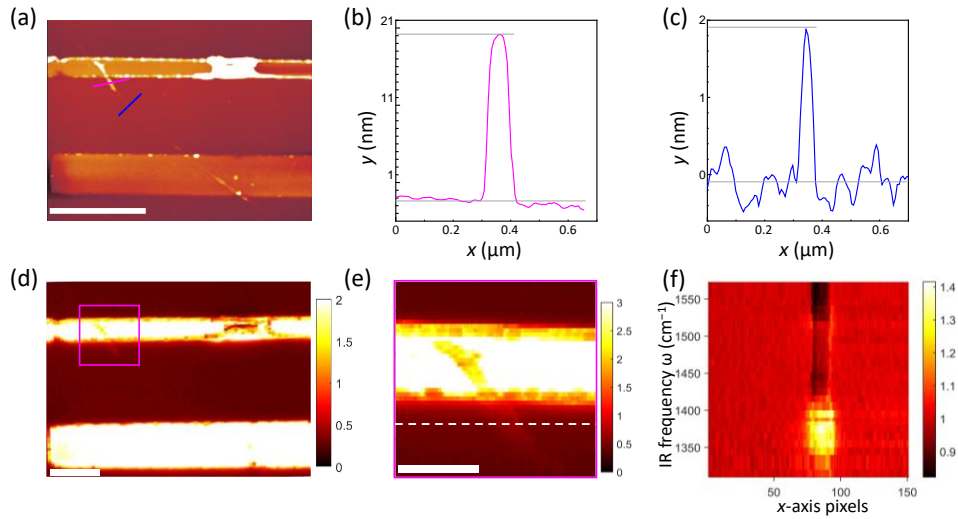

**Figure 2**. (a) AFM topography image of the heterostructure nanotube. Height profiles show that (b) the thick part of the nanotube (purple line in (a)) is about 20 nm while (c) the thin part (blue line in (a)) is around 2 nm. (d) s-SNOM $S_2$-amplitude image of heterostructure nanotube at IR frequency $\omega = 1350$ cm$^{-1}$, the thick part in purple square area is enlarged in (e). (f) s-SNOM hyperspectral cross-section: profiles taken along the dashed white line in (e) and normalized to background, from scans at IR frequencies $1310 \sim 1573$ cm$^{-1}$. Scale bar in (a) 2 μm, (d) 1 μm, and (e) 400 nm.

The prepared 1D heterostructures were then face transferred, taking advantage of a water vapor assisting technique[30], onto the target chip. The SEM images captured after wet transfer present obvious contrasts from some nanotubes while absent from some others as compared in Fig. S4. As characterization results revealed in Fig. 1, the thick and bright parts were successfully covered by BNNT and/or MoS$_2$NT, so we could roughly identify heterogenous parts on specific post-transferred nanotube by SEM and design the metal contacts accordingly to fabricate a possible S-I-S heterojunction diode. Scattering-type scanning near-field optical microscopy (s-SNOM) has been utilized to explore hexagonal boron nitride (hBN)[31-33]. s-SNOM allows to detect the optical material signatures with highest spatial resolution, limited only by its tip radius and signal-to-noise ratio. Here we resolved the 2 nm thin heterostructure nanotube, smeared in a wider pattern by the instrumental tip function in Fig. 2, which was because of a high optical contrast achieved when tuning the excitation laser to a peculiarly strong phonon resonance of hBN. Fig. 2(d) shows the s-SNOM map of one device along with the thick part of the heterostructure nanotube enlarged in Fig. 2(e) at infrared (IR) frequency $\omega = 1350$ cm$^{-1}$. The images were collected in pseudoheterodyne (PsHet) mode of Neaspec s-SNOM (tapping amplitude ~ 70 nm, ARROW-NCPt tips by Nanoworld < 25 nm radius, excitation by quantum cascade laser MIRCat by Daylight in CW mode at power < 2 mW in focal aperture), tuned to show strong s-SNOM signal in all PsHet harmonics $S_1$-$S_4$. $S_2$ was used for Fig. 2 and S5, where a sequence of images at 26 different frequencies is shown, constituting a hyperspectral cube: $S_2(f, x, y)$ for amplitude vs. frequency and two spatial coordinates. Taking a cross-section at the fixed vertical coordinate, $y_o$ (the dashed white line in Fig. 2(e)), reveals spectral dependence of the s-SNOM heterostructure nanotube contrasts as shown in Fig. 2(f). The frequency dependence of amplitude normalized to SiO$_2$ background, $S_2^N(f, x; y_o)$, shows negative contrast (absorption of the excitation light by the nanotube) in the whole spectral region except for a narrow band of positive contrast. This narrow band corresponds to the strongest phonon-polariton resonance of BNNT[34], which happens at frequency of 1370 cm$^{-1}$ and is in consistent with the Fourier transform IR (FTIR) measurement in previous reports[31, 35].

Figure 3 exhibits the diode device we discuss in this work. The SEM image in Fig. 3(a) shows the nanotube before metal contact deposition, and the striking contrasts indicate that the bottom part of the nanotube is thicker than the upper part. Thereupon, the designed metal contacts, which were sputtered with 2 nm Cr and 20 nm Pd into the patterned trenches, are in touch with these two obviously different parts of the nanotube and denoted as electrode E1 and E2, respectively, as shown in Fig. 3(b). Atomic force microscope (AFM) mapping in Fig. 3(c) confirms the previous interpretation and assumption that the contrast appearance along nanotube in SEM images is indeed reflecting the thickness of the nanotube. The height profiles on the two electrodes are illustrated in Fig. 3(d) and 3(e), quantifying the bottom part of the nanotube is as thick as 11 nm while the upper part is as thin as only 1.4 nm, which is a common diameter of SWCNT from our CVD system[36]. Therefore, we can tentatively conclude that electrode E1 is in contact with $MoS_2NT$ while electrode E2 is attached to SWCNT.

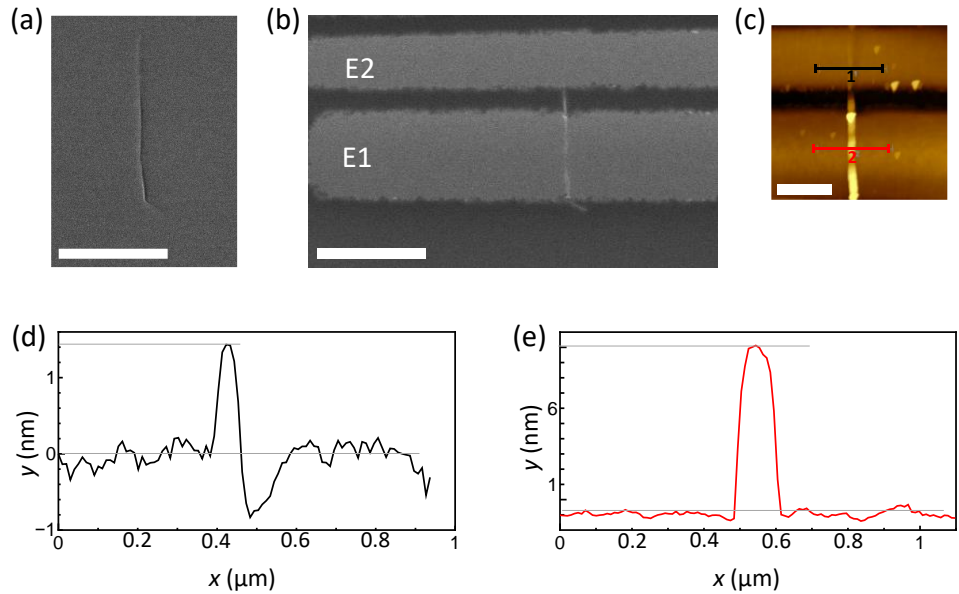

**Figure 3**. SEM images of 1D heterostructure (a) before and (b) after metal contact deposition. (c) is AFM mapping of (b) and the profile 1 and 2 are shown in (d) and (e), respectively. Scale bar in (a) 1 µm, (b) 1 µm, and (c) 500 nm.

Current-voltage (*I-V*) measurements were conducted on the heterostructure device, and a schematic cross-sectional view is depicted in Fig. 4(a). When one electrode is grounded while the other is applied with a driving bias sweeping from −2 V to 2 V, there is a strong rectifying effect no matter the charge carriers were injected from either electrode. The noticeable difference from sweeping voltages on the two electrodes is the voltage polarity of the ON state. As shown in Fig. 4(b), *I-V* curves from the two situations seem centrosymmetric, which is bespeaking a common source for the rectifying effect, but electrode E1 prefers negative voltage polarity while electrode E2 prefers the opposite. To verify that the rectifying effect is originated from the S-I-S heterojunction but not from metal-semiconductor interface, namely Schottky barrier, a normal SWCNT device with the same configuration on the same target chip and transferred from the same as-grown substrate is comparatively analyzed alongside with the heterostructure device as a reference, and it is presented in Fig. S6. First of all, unlike the unevenness of the heterostructure shown in Fig. 3, the referential SWCNT device shows a uniform surface in SEM as can be found in Fig. S6(a) before and (b) after metal deposition. The metal contacts in this work are basically Pd (Cr as an adhesive layer and its thickness of 2 nm is not enough to form an intact thin film), which is believed to provide Ohmic contacts for hole injections to CNT[23], while in the reality, the contact resistance is impeding the smooth transfer of charge carriers, as shown in Fig. S6(c). Electrode E2r is end-bonded with SWCNT and has a larger contact area in comparison with the side-bonded electrode E1r, so the resistance is much smaller when the charge carriers are introduced from electrode E2r, as reflected by the *I-V* curves in Fig. S6(c). Although the contact condition has a strong influence on the *I-V* curve, the rectifying effect deriving from Schottky barrier between metal and semiconducting SWCNT is allowing hole injection and hindering electron injection from both electrodes. Consequently, we can exclude the Schottky barrier as the main origin of the rectifying effect from the heterostructure device in Fig. 4, and there leaves no other factor but the S-I-S heterojunction to account for the effect. The *I-V* curves in the range from −1 V to 1 V of Fig. 4(b) were also finely fitted by a modified diode equation[37] expressed as

$$I = I_S \left( e^{\frac{q(V - I \cdot R_S)}{nk_B T}} - 1 \right) + \frac{V}{R_{sh}} \qquad (\text{Eq. 1})$$

in which, $n$, $q$, $k_B$, $T$ are constants of ideality factor ($n$ = 1.2)[17], electron charge, Boltzman constant, and absolute temperature, respectively. $I_s$, $R_s$, $R_{sh}$ are fitting parameters of dark saturation current, series resistance, and shunt resistance, respectively. Details of the fitting parameters can be found in Table S1.

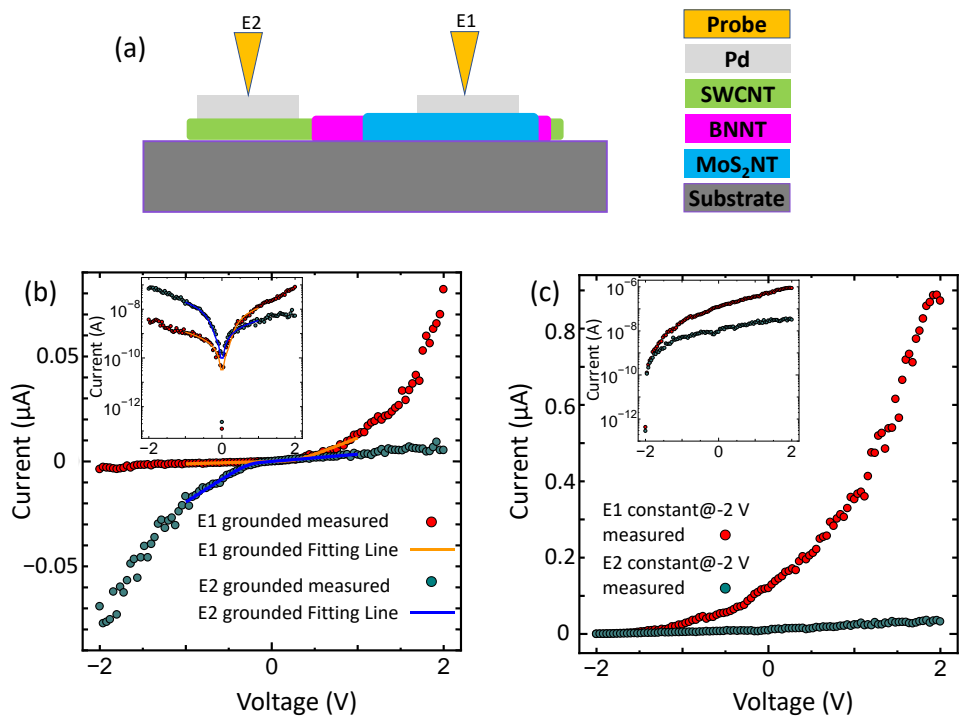

**Figure 4**. (a) Schematic cross-sectional view of the measured 1D heterojunction diode. (b) Electrical characteristics: the red dots are measured when electrode E1 (denoted in the schematic and SEM image of Fig. 3(b)) is grounded while electrode E2 is sweeping from −2 V to 2 V, and the orange solid line is the fitting for experimental data; the olive dots are from the reverse situation and the blue solid line is the fitting for experimental data, the inset shows the same curves in semi-logarithmic axis. (c) Electrical characteristics: the red dots are measured when electrode E1 is applied with a constant voltage of −2 V while electrode E2 is sweeping from −2 V to 2 V, and the olive dots are from the reverse situation, the inset is of the same curves in semi-logarithmic axis.

As analyzed before and illustrated in Fig. 4(a), electrode E1 is in contact with MoS$_2$NT, which is expressing as an *n*-type semiconductor due to natural doping although linked to high work function metal such as Pd, as a result of strong Fermi level pinning[38]; and electrode E2 is covering SWCNT, which shows *p*-type semiconductor characteristics[39]. Therefore, we can expect that a negative voltage polarity on electrode E1 and a positive voltage polarity on electrode E2 will supply a forward bias for this S-I-S heterojunction diode while the reverse scenario will block the flow of current. In Fig. 4(c), one electrode of the heterostructure was applied with a constant voltage of −2 V while the other was driven by a sweeping voltage from −2 V to 2 V. With the forward bias, the current is rapidly increasing with the voltage drop, while the reverse bias results in a much lower current. With a bias drop of 4 V, the rectifying ratio is 24, outperforming previously reported 2D S-I-S diode[20]. On the other side, if a 2D S-I-S diode is as narrow as the 1D heterojunction diode presented here, it will result in degradation of conductance due to disordered edges and will be inadaptable to the need of high-resolution and high-sensitivity photodetectors. Additionally, the same measurements were also performed on the referential SWCNT device as shown in Fig. S6(d). Because both electrodes were biased in these measurements, the effect of voltage barrier caused by contact resistance is indistinguishable no matter which electrode is constantly negatively biased while the driving bias on the other electrode is sweeping. Therefore, *I-V* curves are almost identical when the status of bias is switched between the two contact electrodes. Moreover, as was compared in Reference 20, the existence of insulating BNNT is of crucial importance to guarantee the tunneling transport of charge carriers that leads to higher current and improved rectifying ratio. The current carrying capability is closely related to the band gap of SWCNT channel, therefore, it is impractical to evaluate this merit by comparing the present S-I-S heterojunction diode and normal SWCNT device.

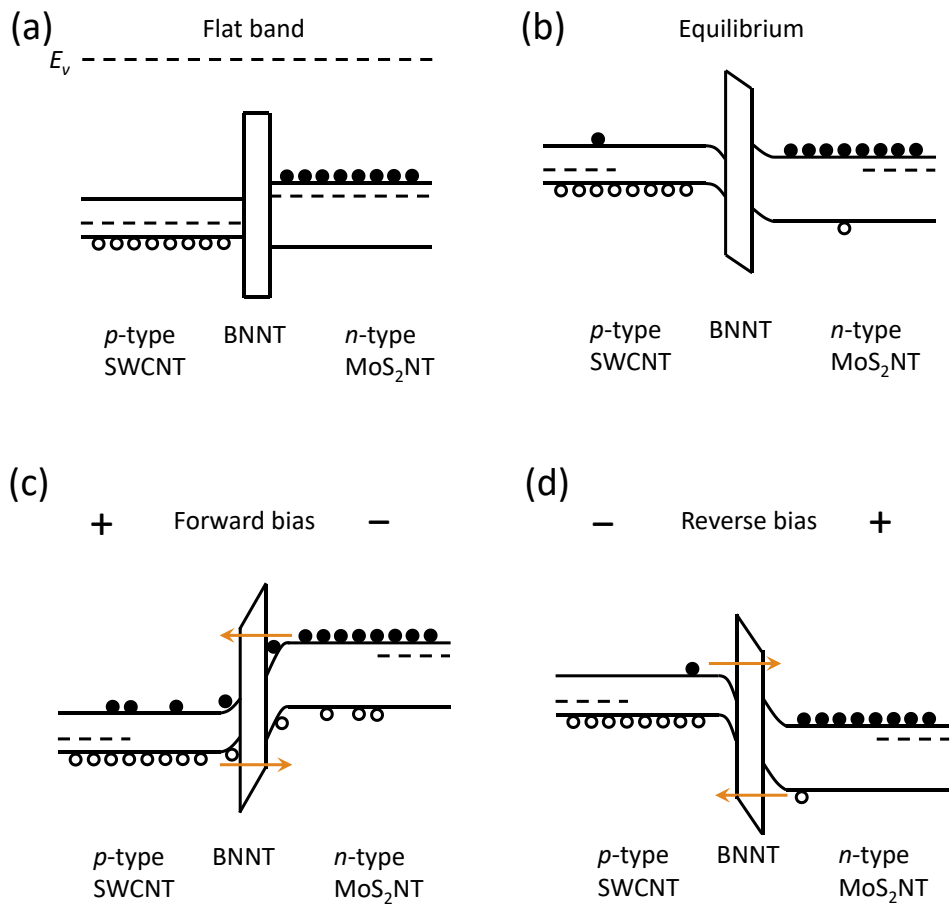

**Figure 5**. Schematics of energy band of 1D S-I-S heterojunction diode: (a) flat band, (b) equilibrium, (c) forward bias, and (d) reverse bias.

The carrier transport mechanism of the S-I-S heterojunction diode is explained in schematics of energy band illustrated in Fig. 5. The band gap of SWCNT is determined by its chirality[40]. As the measured diameter is around 1.4 nm, the chiralities with a close diameter possess a band gap around 0.6 eV[41]. Moreover, the band gap of BNNT is referenced as to be 5.5 eV[42], and we measured photoluminescence (PL) signal from suspended heterostructure nanotube to be 1.88 eV as shown in Fig. S7, close to the previous reports about 2D $MoS_2$[43, 44], so we assigned the contribution of emerging PL signal was from $MoS_2$NT. The work function of SWCNT is 5.05 eV[45], while electron affinities of $MoS_2$NT and BNNT are taken from literature of 4.2 eV[46] and 2 eV[47], respectively. On the basis of these parameters, the flat band of SWCNT/BNNT/$MoS_2$NT sandwich structure is depicted in Fig. 5(a). If the insulating BNNT is very thick, it will stop any carriers transporting and the band diagram will remain flat as in Fig. 5(a). Otherwise, little impediment is met by majority carriers from both sides with a very thin BNNT (monolayer for instance) in between, and an equilibrium status can be reached with small band bending on the edge, exerting a small built-in potential radially on enveloped thick heterostructure nanotube as shown in Fig. 5(b). Besides, the built-in potential will slightly deplete the inner SWCNT, while outside the sandwich structure, exposed SWCNT is still intrinsically *p*-doped, which will make for an electrostatic potential laterally along SWCNT, i.e. lateral $p^-$-*p* junction in series with radial S-I-S heterojunction.

When a forward bias is executed on the heterostructure device, which refers to a positive polarity on SWCNT side while negative polarity on MoS$_2$NT side, as demonstrated in Fig. 5(c), majority carriers from both sides of insulator are rapidly accumulating at the interface driven by potential and flattening the initial band bending if there is; with the increase of bias, the accumulated carriers swiftly tunnel through the insulting layer and lead to a raising current. On the other hand, a reverse bias will build up high potential barrier to immobilize majority carriers and minority carriers can only provide a small current flow as schematically illustrated in Fig. 5(d). This tunneling mechanism provide explanation for the rectifying effect measured in Fig. 4, and the aforementioned lateral $p^-$-$p$ junction in series will additionally contribute to the overall rectification effect. Furthermore, comparing the *I-V* curves of heterojunction diode (in Fig. 4) and normal SWCNT device (in Fig. S6), we can conclude that the contact effect is negligible to the overall performance of the diode, considering the condition of the two contacts are very different in heterojunction diode but no noticeable influence presents when the driving bias switches from one to the other.

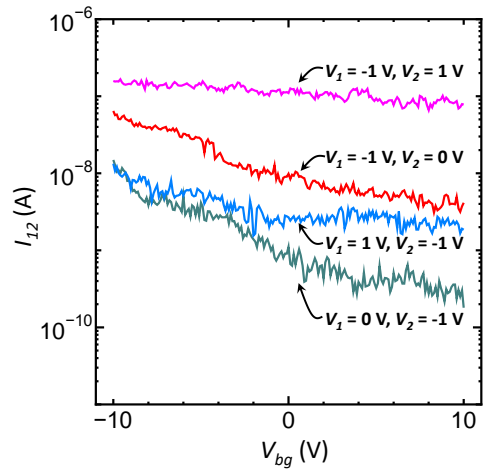

**Figure 6**. Channel current as a function of back gate voltages: magenta solid line is when applied voltage on electrode E1 is −1 V and electrode E2 1 V; red solid line is when applied voltage on electrode E1 is −1 V and electrode E2 0 V; blue solid line is when applied voltage on electrode E1 is 1 V and electrode E2 −1 V; olive solid line is when applied voltage on electrode E1 is 0 V and electrode E2 −1 V.

To further elucidate the underlying working mechanism of the S-I-S heterojunction diode, we performed field effect measurements on the device as shown in Fig. 6. When electrode E1 was supplied with constant voltage −1 V and electrode E2 with 1 V, it was the ON state of the diode, and the current flow between the two electrodes was barely affected by the back gating on silicon which was sweeping from −10 V to 10 V, as the magenta solid curve in Fig. 6 demonstrates. On the other hand, when the device was under OFF state, which was experiencing a reverse bias with electrode E1 of 1 V and electrode E2 of −1 V, the current was more than one magnitude lower than the ON current, as the blue solid curve in Fig. 6 shows, and it starts decreasing slightly with the gating voltage, rendering an improved rectifying ratio. This is analogous to the transfer characteristics of CNT Schottky diode enabled by asymmetric metal contacts[48]. Reducing the driving bias on diode from 2 V to 1 V by decreasing the voltage on electrode E2 from 1 V to 0 V, the ON current was rapidly dropping and the field effect was starting to play a role to tune the ON current as the changes from magenta solid curve to red solid curve imply. The OFF current was experiencing similar changes when voltage on electrode E1 decreased from 1 V to 0 V, but the current drop was smaller than that of the ON current. Since a negative gating voltage was boosting the lower OFF current between two electrodes, we can conclude that the field effect originates from the full channel of SWCNT. Although the performance of *n*-type MoS$_2$ transistor[49] was not as good as *p*-type SWCNT transistor[50] with regard to a unified geometry because of much lower carrier mobility, the field-unaffected ON current presented here indicating a competition from both factors that offset the gating effect, so that the functionality of MoS$_2$NT in the present work was comparable to SWCNT even when it was not serving as the full channel. Therefore, apart from the tremendous potentialities exhibiting by the current 1D S-I-S heterostructure diode in applications such as photodetector and solar cells, after removing the inner SWCNT, the MoS$_2$NT can be a very promising candidate for the next generation electronics or optoelectronics with a steady large bandgap regardless of the chirality. As a good comparison, the transfer characteristics of the normal SWCNT device is presented in Fig. S8.

CONCLUSIONS

Micrometer-long coaxial van der Waals heterostructure nanotube composed of SWCNT, BNNT, and MoS$_2$NT has been synthesized. An S-I-S architecture can be identified given the template SWCNT is semiconducting. Without any intentional interference, semiconducting SWCNT demonstrates *p*-type characteristics with conventional noble metal contacts, while MoS$_2$NT behaves like *n*-type semiconductor. On account of these, we are presenting an S-I-S heterojunction radially which results in a SWCNT-BNNT-MoS$_2$NT diode as narrow as 11 nm. The heterojunction diode demonstrates a rectifying ratio of 24 with a bias voltage of 4 V, which is outperforming a similar 2D heterojunction diode. The insulating BNNT is of crucial importance that not only facilitates the wrapping of MoS$_2$NT and also provides tunneling media that influences ON and OFF current of diode as well as its rectifying ratio, so a precise control growth of heterostructure layers will benefit the performance of such heterostructure devices.

METHODS

**Fabrication processes.** Electron beam lithography (EBL) was utilized to make mark and frame patterns and then the patterns were transferred onto 525-μm-thick Si substrate (100 nm SiO$_2$ on the top) by reactive ion etching (RIE), removing about 200 nm of the top layer. 0.3 nm cobalt (Co) catalysts were deposited by sputtering in EBL patterned areas. Catalyst areas were protected by resist and the exposed areas were etched by RIE and deep reactive ion etching (DRIE) to produce 8-μm-high Si poles. SWCNTs were synthesized bridging Si poles by alcohol catalytic chemical vapor deposition (ACCVD), during which alcohol as carbon source was introduced for 10 minutes at 800 °C. BNNT coating was performed with 30 mg ammonia borane (H$_3$NBH$_3$) as precursor at the upstream and being heated to 70 °C, and sample furnace at 1075 °C for 2 h. MoS$_2$NT coating was followed with sulfur (S) powder at the upstream and being heated to 138 °C, while molybdenum oxide (MoO$_3$) and sample chip keeping 4 cm apart in the furnace at 650 °C for 50 minutes. The as-grown 1D heterostructure nanotubes were face transferred onto target chip assisting by water vapor. Electrical connections were patterned by EBL and realized by sputtering 2 nm chromium (Cr) plus 20 nm palladium (Pd) as metal contacts.

**Characterization of heterostructure nanotubes.** Raman and PL spectra were taken by a Raman spectrometer (inVia, Renishaw) with the excitation wavelength of 532 nm. AES spectrum was obtained through FE-Auger Electron Spectroscopy Model SAM-680. The hyperspectral imaging of heterostructure nanotubes was performed using a customized neaSNOM microscope (Neaspec GmbH): AFM combined with a UV-Vis-NIR-MIR excitation system and electronics. Electrical measurements were conducted in a back-gate geometry in air at room temperature using a semiconductor parameter analyzer (Agilent, 4156C).

ASSOCIATED CONTENT

**Supporting Information.**

Schematics of fabrication processes, SEM images of as-grown and transferred heterostructure nanotubes, full Raman spectrum of suspended heterostructure nanotube, s-SNOM scan images at 26 different IR frequencies, electrical measurements of referential SWCNT device, and PL of suspended heterostructure nanotube.

AUTHOR INFORMATION

**Corresponding Author**

* Shigeo Maruyama Email: maruyama@photon.t.u-tokyo.ac.jp

* Ya Feng Email: fengya@photon.t.u-tokyo.ac.jp

**Author Contributions**

Y.F. and S.M. conceived the project. Y.F. and T.I. developed the device fabrication scheme. Y.F. fabricated silicon-pole chip and synthesized SWCNT. H. L. conducted BNNT and $MoS_2NT$ coating. Y.F. transferred as-grown heterostructure nanotubes and designed diode structure. Y.F. characterized samples by SEM, Raman (PL), and AES. S.V.R. performed s-SNOM measurements. F. Y. took electrical measurements. Y.F., S.V.R. and S.M. analyzed the data. Y.F. wrote the manuscript. All the authors participated in the discussion of the work and commented on the manuscript.

**Notes**

The authors declare no competing financial interest.

ACKNOWLEDGMENT

Part of this work was supported by JSPS KAKENHI Grant Numbers JP18H05329, JP20H00220, and by JST, CREST Grant Number JPMJCR20B5, Japan. Work of S.V.R. was supported by the National Science Foundation (Grant DMR-2011839). Part of this work was conducted at Takeda Sentanchi Supercleanroom, supported by "Nanotechnology Platform" of the Ministry of Education, Culture, Sports, Science and Technology (MEXT), Japan, Grant number JPMXP09F19UT0006.

REFERENCES

1. Sze, S. M.; Ng, K. K., *Physics of Semiconductor Devices*, The Third Edition; John Wiley & Sons, Inc.: Hoboken, New Jersey, 2007.

2. Geim, A. K.; Grigorieva, I. V., Van der Waals heterostructures. *Nature* **2013,** *499*, 419.

3. Liu, Y.; Weiss, N. O.; Duan, X.; Cheng, H.-C.; Huang, Y.; Duan, X., Van der Waals heterostructures and devices. *Nature Reviews Materials* **2016,** *1*, 16042.

4. Lin, Y.-C.; Ghosh, R. K.; Addou, R.; Lu, N.; Eichfeld, S. M.; Zhu, H.; Li, M.-Y.; Peng, X.; Kim, M. J.; Li, L.-J.; Wallace, R. M.; Datta, S.; Robinson, J. A., Atomically thin resonant tunnel diodes built from synthetic van der Waals heterostructures. *Nature Communications* **2015,** *6* (1), 7311.

5. Sarkar, D.; Xie, X.; Liu, W.; Cao, W.; Kang, J.; Gong, Y.; Kraemer, S.; Ajayan, P. M.; Banerjee, K., A subthermionic tunnel field-effect transistor with an atomically thin channel. *Nature* **2015,** *526* (7571), 91-95.

6. Chu, D.; Lee, Y. H.; Kim, E. K., Selective control of electron and hole tunneling in 2D assembly. *Science Advances* **2017,** *3* (4), e1602726.

7. Li, H.-M.; Lee, D.-Y.; Choi, M. S.; Qu, D.; Liu, X.; Ra, C.-H.; Yoo, W. J., Metal-semiconductor barrier modulation for high photoresponse in transition metal dichalcogenide field effect transistors. *Scientific Reports* **2014,** *4* (1), 4041.

8. Lee, C.-H.; Lee, G.-H.; van der Zande, A. M.; Chen, W.; Li, Y.; Han, M.; Cui, X.; Arefe, G.; Nuckolls, C.; Heinz, T. F.; Guo, J.; Hone, J.; Kim, P., Atomically thin p–n junctions with van der Waals heterointerfaces. *Nature Nanotechnology* **2014,** *9* (9), 676-681.

9. Withers, F.; Del Pozo-Zamudio, O.; Mishchenko, A.; Rooney, A. P.; Gholinia, A.; Watanabe, K.; Taniguchi, T.; Haigh, S. J.; Geim, A. K.; Tartakovskii, A. I.; Novoselov, K. S., Light-emitting diodes by band-structure engineering in van der Waals heterostructures. *Nature Materials* **2015,** *14* (3), 301-306.

10. Pyatkov, F.; Fütterling, V.; Khasminskaya, S.; Flavel, B. S.; Hennrich, F.; Kappes, M. M.; Krupke, R.; Pernice, W. H. P., Cavity-enhanced light emission from electrically driven carbon nanotubes. *Nature Photonics* **2016,** *10*, 420.

11. Qiu, C.; Zhang, Z.; Xiao, M.; Yang, Y.; Zhong, D.; Peng, L.-M., Scaling carbon nanotube complementary transistors to 5-nm gate lengths. *Science* **2017,** *355* (6322), 271-276.

12. Graf, A.; Held, M.; Zakharko, Y.; Tropf, L.; Gather, M. C.; Zaumseil, J., Electrical pumping and tuning of exciton-polaritons in carbon nanotube microcavities. *Nature Materials* **2017,** *16*, 911.

13. Cui, K.; Anisimov, A. S.; Chiba, T.; Fujii, S.; Kataura, H.; Nasibulin, A. G.; Chiashi, S.; Kauppinen, E. I.; Maruyama, S., Air-stable high-efficiency solar cells with dry-transferred single-walled carbon nanotube films. *Journal of Materials Chemistry A* **2014,** *2* (29), 11311-11318.

14. Zhou, C.; Kong, J.; Yenilmez, E.; Dai, H., Modulated chemical doping of individual carbon nanotubes. *Science* **2000,** 290 (5496), 1552-1555.

15. Kang, B. R.; Yu, W. J.; Kim, K. K.; Park, H. K.; Kim, S. M.; Park, Y.; Kim, G.; Shin, H.-J.; Kim, U. J.; Lee, E.-H.; Choi, J.-Y.; Lee, Y. H., Restorable type conversion of carbon nanotube transistor using pyrolytically controlled antioxidizing photosynthesis coenzyme. *Advanced Functional Materials* **2009,** 19 (16), 2553-2559.

16. Peng, L.-M.; Zhang, Z.; Wang, S.; Liang, X., A doping-free approach to carbon nanotube electronics and optoelectronics. *AIP Advances* **2012,** *2* (4), 041403.

17. Lee, J. U.; Gipp, P. P.; Heller, C. M., Carbon nanotube p-n junction diodes. *Applied Physics Letters* **2004,** *85* (1), 145-147.

18. Mueller, T.; Kinoshita, M.; Steiner, M.; Perebeinos, V.; Bol, A. A.; Farmer, D. B.; Avouris, P., Efficient narrow-band light emission from a single carbon nanotube p–n diode. *Nature Nanotechnology* **2009,** *5*, 27.

19. Gupta, G.; Rajasekharan, B.; Hueting, R. J. E., Electrostatic doping in semiconductor devices. *IEEE Transactions on Electron Devices* **2017,** *64* (8), 3044-3055.

20. Jeong, H.; Bang, S.; Oh, H. M.; Jeong, H. J.; An, S.-J.; Han, G. H.; Kim, H.; Kim, K. K.; Park, J. C.; Lee, Y. H.; Lerondel, G.; Jeong, M. S., Semiconductor–insulator–semiconductor diode consisting of monolayer $MoS_2$, h-BN, and GaN heterostructure. *ACS Nano* **2015,** *9* (10), 10032-10038.

21. Wang, L.; Meric, I.; Huang, P.; Gao, Q.; Gao, Y.; Tran, H.; Taniguchi, T.; Watanabe, K.; Campos, L.; Muller, D., One-dimensional electrical contact to a two-dimensional material. *Science* **2013,** *342* (6158), 614-617.

22. Xiang, R.; Inoue, T.; Zheng, Y.; Kumamoto, A.; Qian, Y.; Sato, Y.; Liu, M.; Tang, D.; Gokhale, D.; Guo, J.; Hisama, K.; Yotsumoto, S.; Ogamoto, T.; Arai, H.; Kobayashi, Y.; Zhang, H.; Hou, B.; Anisimov, A.; Maruyama, M.; Miyata, Y.; Okada, S.; Chiashi, S.; Li, Y.; Kong, J.; Kauppinen, E. I.; Ikuhara, Y.; Suenaga, K.; Maruyama, S., One-dimensional van der Waals heterostructures. *Science* **2020,** *367* (6477), 537.

23. Javey, A.; Guo, J.; Wang, Q.; Lundstrom, M.; Dai, H., Ballistic carbon nanotube field-effect transistors. *Nature* **2003,** *424* (6949), 654-657.

24. Fathipour, S.; Remskar, M.; Varlec, A.; Ajoy, A.; Yan, R.; Vishwanath, S.; Rouvimov, S.; Hwang, W. S.; Xing, H. G.; Jena, D.; Seabaugh, A., Synthesized multiwall $MoS_2$ nanotube and nanoribbon field-effect transistors. *Applied Physics Letters* **2015,** 106 (2), 022114.

25. Léonard, F.; Stewart, D. A., Properties of short channel ballistic carbon nanotube transistors with Ohmic contacts. *Nanotechnology* **2006,** *17* (18), 4699-4705.

26. Franklin, A. D.; Koswatta, S. O.; Farmer, D. B.; Smith, J. T.; Gignac, L.; Breslin, C. M.; Han, S.-J.; Tulevski, G. S.; Miyazoe, H.; Haensch, W.; Tersoff, J., Carbon nanotube complementary wrap-gate transistors. *Nano Letters* **2013,** *13* (6), 2490-2495.

27. Chen, X.; Roemer, M.; Yuan, L.; Du, W.; Thompson, D.; del Barco, E.; Nijhuis, C. A., Molecular diodes with rectification ratios exceeding $10^5$ driven by electrostatic interactions. *Nature Nanotechnology* **2017,** *12* (8), 797-803.

28. Capozzi, B.; Xia, J.; Adak, O.; Dell, E. J.; Liu, Z.-F.; Taylor, J. C.; Neaton, J. B.; Campos, L. M.; Venkataraman, L., Single-molecule diodes with high rectification ratios through environmental control. *Nature Nanotechnology* **2015,** *10* (6), 522-527.

29. Arai, H.; Inoue, T.; Xiang, R.; Maruyama, S.; Chiashi, S., Non-catalytic heteroepitaxial growth of aligned, large-sized hexagonal boron nitride single-crystals on graphite. *Nanoscale* **2020,** *12* (18), 10399-10406.

30. Abrams, Z. e. R.; Ioffe, Z.; Tsukernik, A.; Cheshnovsky, O.; Hanein, Y. J. N. l., A complete scheme for creating predefined networks of individual carbon nanotubes. *Nano Letters* **2007,** *7* (9), 2666-2671.

31. Dai, S.; Fei, Z.; Ma, Q.; Rodin, A. S.; Wagner, M.; McLeod, A. S.; Liu, M. K.; Gannett, W.; Regan, W.; Watanabe, K.; Taniguchi, T.; Thiemens, M.; Dominguez, G.; Neto, A. H. C.; Zettl, A.; Keilmann, F.; Jarillo-Herrero, P.; Fogler, M. M.; Basov, D. N., Tunable phonon polaritons in atomically thin van der Waals crystals of boron nitride. *Science* **2014,** *343* (6175), 1125-1129.

32. Woessner, A.; Lundeberg, M. B.; Gao, Y.; Principi, A.; Alonso-González, P.; Carrega, M.; Watanabe, K.; Taniguchi, T.; Vignale, G.; Polini, M.; Hone, J.; Hillenbrand, R.; Koppens, F. H. L., Highly confined low-loss plasmons in graphene–boron nitride heterostructures. *Nature Materials* **2015,** *14* (4), 421-425.

33. Dai, S.; Quan, J.; Hu, G.; Qiu, C.-W.; Tao, T. H.; Li, X.; Alù, A., Hyperbolic phonon polaritons in suspended hexagonal boron nitride. *Nano Letters* **2019,** *19* (2), 1009-1014.

34. Xu, X. G.; Ghamsari, B. G.; Jiang, J.-H.; Gilburd, L.; Andreev, G. O.; Zhi, C.; Bando, Y.; Golberg, D.; Berini, P.; Walker, G. C., One-dimensional surface phonon polaritons in boron nitride nanotubes. *Nature Communications* **2014,** *5* (1), 4782.

35. Wang, P.; Zheng, Y.; Inoue, T.; Xiang, R.; Shawky, A.; Watanabe, M.; Anisimov, A.; Kauppinen, E. I.; Chiashi, S.; Maruyama, S., Enhanced in-plane thermal conductance of thin films composed of coaxially combined single-walled carbon nanotubes and boron nitride nanotubes. *ACS Nano* **2020,** *14* (4), 4298-4305.

36. Feng, Y.; Inoue, T.; An, H.; Xiang, R.; Chiashi, S.; Maruyama, S., Quantitative study of bundle size effect on thermal conductivity of single-walled carbon nanotubes. *Applied Physics Letters* **2018,** *112* (19), 191904.

37. Ho, X.; Ye, L.; Rotkin, S. V.; Xie, X.; Du, F.; Dunham, S.; Zaumseil, J.; Rogers, J. A., Theoretical and experimental studies of Schottky diodes that use aligned arrays of single-walled carbon nanotubes. *Nano Research* **2010,** *3* (6), 444-451.

38. Kim, C.; Moon, I.; Lee, D.; Choi, M. S.; Ahmed, F.; Nam, S.; Cho, Y.; Shin, H.-J.; Park, S.; Yoo, W. J., Fermi level pinning at electrical metal contacts of monolayer molybdenum dichalcogenides. *ACS Nano* **2017,** *11* (2), 1588-1596.

39. Franklin, A. D.; Farmer, D. B.; Haensch, W., Defining and overcoming the contact resistance challenge in scaled carbon nanotube transistors. *ACS Nano* **2014,** *8* (7), 7333-7339.

40. Dresselhaus, M.; Eklund, P., Phonons in carbon nanotubes. *Advances in Physics* **2000,** *49* (6), 705-814.

41. Wilder, J. W. G.; Venema, L. C.; Rinzler, A. G.; Smalley, R. E.; Dekker, C., Electronic structure of atomically resolved carbon nanotubes. *Nature* **1998,** *391* (6662), 59-62.

42. Jaffrennou, P.; Barjon, J.; Lauret, J.-S.; Maguer, A.; Golberg, D.; Attal-Trétout, B.; Ducastelle, F.; Loiseau, A., Optical properties of multiwall boron nitride nanotubes. *Physica Status Solidi (B)* **2007,** *244* (11), 4147-4151.

43. Splendiani, A.; Sun, L.; Zhang, Y.; Li, T.; Kim, J.; Chim, C.-Y.; Galli, G.; Wang, F., Emerging photoluminescence in monolayer $MoS_2$. *Nano Letters* **2010,** *10* (4), 1271-1275.

44. Choi, M. S.; Lee, G.-H.; Yu, Y.-J.; Lee, D.-Y.; Lee, S. H.; Kim, P.; Hone, J.; Yoo, W. J., Controlled charge trapping by molybdenum disulphide and graphene in ultrathin heterostructured memory devices. *Nature Communications* **2013,** *4* (1), 1624.

45. Shiraishi, M.; Ata, M., Work function of carbon nanotubes. *Carbon* **2001,** *39* (12), 1913-1917.

46. Schlaf, R.; Lang, O.; Pettenkofer, C.; Jaegermann, W., Band lineup of layered semiconductor heterointerfaces prepared by van der Waals epitaxy: Charge transfer correction term for the electron affinity rule. *Journal of Applied Physics* **1999,** *85* (5), 2732-2753.

47. Cumings, J.; Zettl, A., Field emission and current-voltage properties of boron nitride nanotubes. *Solid State Communications* **2004,** *129* (10), 661-664.

48. Yang, M. H.; Teo, K. B. K.; Milne, W. I.; Hasko, D. G., Carbon nanotube Schottky diode and directionally dependent field-effect transistor using asymmetrical contacts. *Applied Physics Letters* **2005,** *87* (25), 253116.

49. Radisavljevic, B.; Radenovic, A.; Brivio, J.; Giacometti, V.; Kis, A., Single-layer MoS$_2$ transistors. *Nature Nanotechnology* **2011,** 6 (3), 147-150.

50. Franklin, A. D.; Luisier, M.; Han, S.-J.; Tulevski, G.; Breslin, C. M.; Gignac, L.; Lundstrom, M. S.; Haensch, W. J. N. l., Sub-10 nm carbon nanotube transistor. *Nano letters* **2012,** *12* (2), 758-762.